\documentclass{rspublic}

\usepackage{graphicx}

\newcommand{\be}{\begin{equation}}
\newcommand{\ee}{\end{equation}}
\newcommand{\vxi}{{\mathbf \xi}}

\newcommand{\x}{{\mathbf x}}

\newcommand{\X}{{\mathbf X}}

\newcommand{\opA}{{\widehat{A}}}

\newcommand{\opR}{{\widehat{R}}}

\newcommand{\oprho}{{\widehat{\rho}}}

\begin{document}

\title{Husimi-Wigner representation of chaotic eigenstates}

\author[F. Toscano, A. Kenfack, A. R. R. Carvalho, J. M. Rost, A. M. Ozorio de Almeida]
{Fabricio Toscano$^{1,2}$, Anatole Kenfack$^3$, Andre R. R. Carvalho$^4$, 
Jan~M.\ Rost$^3$, Alfredo M. Ozorio de Almeida$^5$\footnote{ozorio@cbpf.br}}

\affiliation{
$^1$
Funda\c c\~ao Centro de Ci\^encias e Educa\c c\~ao Superior a Dist\^ancia do Estado do Rio de Janeiro, 20943-001 Rio de Janeiro, RJ, Brazil.\\
$^2$
Instituto de F\'{\i}sica, Universidade Federal do Rio de Janeiro, Cx. P. 68528, 21941-972 Rio de Janeiro, RJ, Brazil.\\
$^3$ Max-Planck-Institut f\"ur Physik Komplexer Systeme, N\"othnitzer Strasse 38, D-01187 Dresden, Germany\\
$^4$ Australian Natl Univ, Fac Sci, Dept Phys, Canberra, ACT 0200 Australia\\
$^5$Centro Brasileiro de Pesquisas Fisicas, Rua Xavier Sigaud 150, 22290-180 Rio de Janeiro, R.J., Brazil}

\label{firstpage}

\maketitle

\begin{abstract}{phase space representations, Wigner function, Husimi function, semiclassical mechanics, chaotic eigenstates.}

Just as a coherent state may be considered as a quantum point,
its restriction to a factor space of the full Hilbert space
can be interpreted as a quantum plane. 
The overlap of such a factor coherent state 
with a full pure state is akin to a quantum section.
It defines a reduced pure state in the cofactor Hilbert space.
The collection of all the Wigner functions corresponding to a full set
of parallel quantum sections defines the Husimi-Wigner reresentation.
It occupies an intermediate ground 
between drastic suppression of nonclassical features,
characteristic of Husimi functions, 
and the daunting complexity of higher dimensional Wigner functions.
After analysing these features for simpler states, 
we exploit this new representation as a probe
of numerically computed eigenstates of chaotic Hamiltonians.
The individual two-dimensional Wigner functions 
resemble those of semiclassically quantized states,
but the regular ring pattern is broken by dislocations.  

\end{abstract}

\section{Introduction}

It is well known that phase space representations of quantum mechanics
are powerful tools for studying the correspondence between the density
operator and classical distributions in phase space. 
The several choices of representation are partially distinguished
by the different ways they highlight classical structures against
a background of quantum interferences. In the case of the Wigner function
(Wigner  1932), $W(\x)$, in terms of the phase space variables, 
$\x=(\x_1,...\x_L)=(p_1,...p_L, q_1,...,q_L)$,
the oscillations due to interferences 
may even have higher amplitudes than the classical region.
In contrast, the Husimi function (Husimi 1940, Takashi  1986), 
which may be defined as a coarse-graining 
of the Wigner function, 
\footnote{Here and throughout, we make the convenient choice
that the frequency of the harmonic oscilator 
for the coherent states is $\omega=m=1$.}
\begin{equation}
\mathcal{H}({\X})= (\pi \hbar)^{-L} \int {\rm d}\x\; 
W(\x)\;\; \exp {\frac{-(\x-\X)^2}{\hbar}},
\label{coarse}
\end{equation}
subtly disguises information on quantum coherences
to the point that they may be numerically undetectable,
while clearly displaying most classical structures (Toscano \& Ozorio de Almeida 1999).

In the case of ($2L$)-dimensional phase spaces ($2L$-D), with $L>1$,
the approximate classical support for a quantum state may take the form
of a discrete set of points. These correspond to either a coherent state,
squeezed or not, or their supperposition, sometimes known as 
{\it Schr\"odinger cat states}. 
Alternatively, semiclassical {\it Van Vleck states} (Van Vleck 1999)
correspond to $L$-D (Lagrangian) surfaces. Of even higher dimension is the
support of {\it ergodic states} satisfying Schnirelman's theorem
 (Shnirelman 1974, Colin de Verdi\`ere 1985 \&  Zelditch 1987):
eigenstates of (classically) chaotic Hamitonians, 
supported by the full $(2L-1)$-D energy shell.
Of course, all these types of state can be superposed in various ways,
which in their turn produce new interferences.

Even though we cannot directly visualize such classical structures 
in a higher dimensional phase space, they will show up in 
appropriate 2-D sections of the corresponding Husimi function.
Even so, it will be virtually impossible to extricate
the crucial quantum phase information in this representation, 
unless all analytical properties concerning the state are known 
(Leboeuf \& Voros  1990 and Leboeuf  \& Voros  1995).
The situation for the Wigner function is just the opposite:
all phase information is immediately available in the oscillatory interference pattern, 
but it becomes hard to sort out its embarassing richness.
For example, a 2-D section may contain a (plain) periodic orbit.
States that are {\it scarred} by this periodic orbit 
(Heller 1984,  Bogomolny 1988 \& Berry  1989 )
are clearly distinguishable in the section of the
Wigner function through this plane (Toscano \textit{et al.}  2001), however, interferences
also arise which can only be generated by classical structures
that are nowhere near this 2-D plane. There is certainly need
for a more manageable representation of the interference effects 
that decorate classical structures of general pure states. 
Indeed, the Wigner function itself,
in the simple case that $L=1$, is a good example of 
a comprehensible interference pattern (Berry 1977a).

Our purpose here is to develop a new tool for the analysis of the
chaotic eigenstates of higher dimensional systems. Even though
there has been continuing interest in their localization and
statistical properties, starting with references Voros 1976 and Berry 1977b,
subsequent work in this field has concentrated on the description
of chaotic eigenstates of quantum maps 
(Hannay 1998, Leboeuf \& Voros  1990 and Leboeuf  \& Voros  1995)
and Schanz  2005 (which contains many further references in this field).
The reason for this is precisely to avoid the difficulties 
of coping with higher dimensions.
The detailed analysis of ergodic states for $L>1$ is
still in its preliminary stages. Their Husimi functions
should be  concentrated in a narrow neighbourhood of 
the energy shell, but the often repeated statement, that this also holds
for the Wigner function, is false. Indeed, it has been shown by Ozorio de Almeida  
\textit{et al.}  2004 that all large scale pure states must have correspondingly small scale
oscillations in their Wigner functions. Such {\it subplanckian structures}
(Zurek 2001) do not contribute to the averages of smooth observables
(hence, the ergodicity over the energy shell). However, future refinements
of experimental techniques will inevitably lead to measurable interference effects.
We here introduce a conceptual tool with which to
probe into the details of these higher dimensional states.

The joint Husimi-Wigner representation of quantum mechanics,
{\bf huwi representation} for short, 
avoids the extremes of near-classicality, 
or of excessive quantum complexity, 
that characterize alternatively its parent representations. 
Decomposing the phase space variables in the Wigner function, $W(\x)$, as
$\x=(\x_1, \x')$, with $\x'=(\x_2,...,\x_L)$ 
and specifying a 2-D plane by the $2(L-1)$ equations, $\x'= X'=const.$,
the huwi function is defined as
\be
h\!\!w_{\X'}(\x_1)= (\pi \hbar)^{-(L-1)} \int {\rm d}\x'\; 
 W(\x_1, \x')\;\; \exp {\frac{-(\x'-\X')^2}{\hbar}}.
\label{huwi1}
\ee
In the following section, the general features of this complete representation
of the density operator in the phase space, $(\x_1, \X')$ will be discussed.
However, our main focus will be centred on the huwi function for a fixed $\X'$.
Viewed classically, this would be merely a slight thickening
of a plane section of the Wigner function, 
but it is a truly quantum section, 
that is, for each parameter, $\X'$, 
$h\!\!w_{\X'}(\x_1)$ represents a different pure state. 
All of these states belong to the same factor Hilbert space
which corresponds to the phase plane, $\x_1$.
This general scenario is developed in the following section.

In  \S 3, we discuss the huwi representation in the simple case of 
coherent states and their superposition. This already exemplifies the
convenient way in which the thickened section erases, 
not only the classical regions foreign to the section, 
but also all their interference effects. 
This is also the clearest setting in which to discuss rotations
and other classical canonical transformations 
as tools to bring desired features into view. 
Van Vleck states are then analyzed in \S 4:
It is shown that generically their huwi representation  
can be approximated by Gaussians in the $\x_1$ phase plane, 
corresponding to (squeezed) coherent states, 
or generalized Schr\"odinger cat states,
centred on the discrete set of points 
where the constant $\X'$-plane intersects the classical surface.

Ergodic states have so far evaded any compact analytical characterization
in any representation, 
so their study in \S 5 must relie on computational evidence.
Given that the 2-D section of a compact $(2L-1)$-D energy shell
is a closed curve, we investigate the family resemblence between $h\!\!w_{\X'}(\x_1)$
and the eigenstates of a Hamiltonian,
defined so that its level curve coincides with the section 
of the higher dimensional energy shell. 
The computational huwi patterns discussed in \S 6 
suggest that these new kinds of pure states
are characterized by dislocations in their 2-D Wigner functions that are
similar to those found in the wave trains of short pulses (Nye \& Berry 1974).

\section{Quantum sections}

Coherent states, $|\X\rangle$, defined in the position representation as
\begin{eqnarray}
\langle q|\X\rangle= \Big(\frac{1}{\pi \hbar}\Big)^{L/4}\exp\Big(-\frac{1}
{2\hbar}(q-Q)^2+\frac{i}{\hbar}P \cdot(q-\frac{Q}{2})\Big),
\label{coherent}
\end{eqnarray}
form a basis for Hilbert space that is overcomplete 
(Glauber 1963, Sudarshan 1963, Klauder \& Skagerstam 1985, Perelemov 1986, 
Schleich 2001 and Cohen Tannoudji \textit{et al.} 1977):
The decomposition,
\begin{eqnarray}
|\psi\rangle=\frac{1}{(2\pi\hbar)^L}\int \rm d\X \; |\X\rangle \langle \X|\psi\rangle,
\end{eqnarray} 
is unique. Considering the full Hilbert space
as a tensor product of factor spaces, 
${\bf H}={\bf H}_1\otimes...{\bf H}_l\otimes...{\bf H}_L$,
for each of the $L$ degrees of freedom,
it is obvious that the coherent state basis also factors, i.e.,
\be
|\X\rangle=|\X_1\rangle \otimes...|\X_l\rangle \otimes...|\X_L\rangle.
\ee
Therefore, decomposing again $\X=(\X_1, \X')$ and taking the partial overlaps, 
\be
|\X_1\rangle = \langle \X'|\X\rangle = |\X_1\rangle \langle {\X}_2|\X_2\rangle...\langle {\X}_L|\X_L\rangle
\ee
generates a basis for the factor space, ${\bf H}_1$, 
such that each $|\psi_{\X'}\rangle=\langle \X'|\psi\rangle$
is a pure state with its wave function, 
\be
\langle \X'|\psi\rangle(q_1)= \int {\rm d}q' \;\langle q'|\X'\rangle^* \;\langle q_1,q'|\psi\rangle.  
\label{sectwave}
\ee

The Husimi function can be defined directly in terms of 
coherent states, alternatively to (\ref{coarse}). 
Given the density operator for a pure state as $\oprho_\psi=|\psi\rangle\langle\psi|$, then
\be
\mathcal{H}_\psi(\X)= \langle \X|\rho|\X\rangle = 
{\rm tr}\>\oprho_\psi\>|\X\rangle\langle \X|=|\langle \X|\psi\rangle|^2 .
\ee
This diagonal coherent state representation also contains full information
concerning the density operator, 
because of the overcompleteness of the coherent state basis.
The alternative definition (\ref{coarse}) of the Husimi function
is an immediate consequence of (Gr\"oenewold  1946, Ozorio de Almeida 1998)
\begin{equation}
\rm tr\> \widehat{\rho}_2\widehat{\rho}_1 = (2\pi\hbar )^L \int d{\x}\; W_2 (\x) W_1 (\x).
\label{traceprod}
\end{equation}

The partial overlap of both the bra and the ket of
$\oprho=|\psi\rangle\langle\psi|$ 
with the same factor coherent state, $|\X'\rangle$,
defines a reduced density operator, 
\be
\oprho_{\X'}= \langle \X'|\rho|\X'\rangle,
\label{sandw}
\ee
in the factor Hilbert space ${\bf H_1}$.
We also recall that the Wigner function in the full phase space 
is defined as (Royer 1977, Ozorio de Almeida 1998)
\be
W(\x)= (\pi\hbar )^{-L} {\rm tr}\>\oprho\> \opR_{\x},
\ee
where the operator for the reflection through the point $\x$ is
\begin{equation}
\label{quantrefleq}
\hat{R}_\x = 2^{-L}  \int \rm d Q 
      \left| \mathbf q + \frac{Q}{2} \right\rangle 
\left\langle \mathbf q - \frac{Q}{2} \right|
       e^{i \mathbf p \cdot Q / \hbar}
\end{equation}
and that similar definitions hold for Wigner functions defined in subspaces, 
with the appropriate adaptation of notation.
Then we find that the alternative definition for the huwi function, 
\be
h\!\!w_{\X'}(\x_1)= (\pi\hbar )^{-1} {\rm tr}\>\oprho_{\X'}\> \opR_{\x_1}
=(\pi\hbar )^{-1} {\rm tr}\>\oprho\> [\opR_{\x_1}\otimes|\X'\rangle\langle \X'|],
\label{huwi2}
\ee 
is clearly equivalent to (\ref{huwi1}).  
One can also reinterpret (\ref{huwi2}) as the Wigner function that represents
the pure state, $\oprho_1={\rm tr'}\oprho\>|\X'\rangle\langle \X'|$,
where the trace is over the factor Hilbert space corresponding to the $X'$ coordinates.
If this state is represented by its Husimi function, then
\be
\mathcal{H}_{X'}(\x_1)=\mathcal{H}(\x_1, \X'),
\label{Hsection}
\ee  
that is, the Husimi function for this quantum section is just the 
section of the full Husimi function. This should not be confused
with the {\it Quantum Poincare Surface of Section} 
(Leboeuf   \& Saraceno 1990a  and Leboeuf   \&  Saraceno 1990b),
which will be discussed in \S 5. 

The definition of the huwi function is based on a wide, but rarely used
freedom in the choice of representations of tensor products of Hilbert spaces.
These correspond classically to Cartesian products of phase planes
and it is more usual to exploit alternative generating functions
for canonical transformations by exchanging variables in classical mechanics 
(Arnold 1978, Goldstein 1980).
However, the corresponding matrix elements in quantum mechanics,
\be
\langle q_1, p_2,...|\opA|{p'}_1, {p'}_2,...\rangle
= {\rm tr}\>\opA\;[|{p'}_1\rangle\langle q_1|\otimes|{p'}_2\rangle\langle p_2|\otimes...],
\ee
also form a faithfull representation of the operator $\opA$, 
for all choices of either $p_j$, or $q_j$, and of ${p'}_j$, or ${q'}_j$.
Furthermore, it is possible to include operators $\opR_{\x_j}$ in the same class as the
dyadic operators $|{p'}_j\rangle\langle q_j|$ as a faithfull basis for 
representing operators. Indeed they may be interpreted as merely defining
alternative planes in the doubled phase space that corresponds to operators, 
just as ordinary phase space corresponds to Hilbert space
(see e.g. Ozorio de Almeida 2007). 
Alternatively, the Husimi basis, $|\X_j\rangle\langle \X_j|$, can also 
be used for any of the degrees of freedom.
In the case of the classical generating functions, switches of representation
are usually motivated by the need to avoid singularities, i.e. caustics,
in the implicit definition of the transformation. These are also a problem
for semiclassical aproximations to quantum evolution.
In the present context, the huwi representation
is singled out by the clarity with which it exhibits both classical and quantum
characteristics of a pure state.

The reduced density operator, $\oprho_{\X'}= \langle \X'|\oprho|\X'\rangle$,
resembles in some respects the partial trace, $\oprho_1={\rm tr'}\oprho$, which is also
a density operator. The corresponding Wigner function is just (Ozorio de Almeida 2007),
\be
W_1(\x_1) =(\pi\hbar )^{-1} {\rm tr}\>\oprho\> [\opR_{\x_1}\otimes {\widehat I}']
=  \int {\rm d}\x'\; W(\x_1, \x'),
\ee
where ${\widehat I}$ is the identity operator. 
However, $\oprho_1$ will not be a pure state unless $\oprho$ can be
factored into product states, whereas $\oprho_{\X'}$ will only be a mixed state
if $\oprho$ is not itself a pure state in the full Hilbert space.
The Fourier transform of $W_1(\x_1)$ is the chord function, 
or the quantum characteristic function for $\oprho_1$:
\be
\chi_1({\bf \xi}_1) = \frac{1}{2\pi\hbar} \int 
\x_1 
\exp{\left(-\frac{i}{\hbar}\vxi_1\wedge\x_1\right)} W_1(\x_1)
\ee
(where the skew product in the exponent is here just a plane vector product, a scalar).
This can be obtained directly from the chord function, $\chi(\vxi)$ for $\oprho$ as
(Ozorio de Almeida 2007)
\be
\chi_1(\vxi_1)=(2\pi\hbar)^{2(L-1)}\>\chi(\vxi_1,\vxi'\!=\! 0),
\label{redchord}
\ee
that is, as a mere section of the full chord function.
On the other hand, it is easy to see that the chord function for $\oprho_{\X'}$,
i.e. the Fourier transform of $h\!\!w_{\X'}(\x_1)$ is
\be
\chi_{\X'}(\vxi_1)= (\pi \hbar)^{-(L-1)} \int {\rm d}\vxi'\; 
 \chi(\vxi_1, \vxi')\;\; \exp {\frac{-(\vxi')^2- i \X' \wedge \vxi'}{\hbar}}.
\ee
Thus, both the chord and the Wigner representations of $\oprho_{\X'}$ are obtained
by integrating over the respective representation of $\oprho$ with a Gaussian window
of width $\sqrt {\hbar}$. In contrast, we may picture these windows for 
the partial trace, $\oprho_1$, as infinitely wide, in the case of the Wigner
function and infinitely narrow, in the case of the chord function.

If the projection of $W(\x)$ is carried out over the momentum half space, $p$,
instead of the components, $\x'$, we obtain the wave intensity, 
$|\langle q|\psi\rangle|^2$, for a pure state. Similar probability densities for linear
combinations of position and momentum result from projections 
in other phase space directions, because of the invariance of 
the Wigner function with respect to linear canonical transformations,
i.e. symplectic transformations. Indeed, at least for $L=1$,  
it is possible to regenerate the Wigner function 
from a subgroup of symplectically related projections
through the Radon transform (Deans 1983).

A point of practical importance concerns normalization. 
The definition (\ref{sandw}) should  be divided by
$N(\X')={\rm tr_1\;\oprho_{\X'}}$, so as to represent a normalized
density operator. Thus, the integral of $h\!\!w_{\X'}(\x_1)$ over all $\x_1$
is $N(\X')$ and the corresponding chord function 
must have $\chi_{\X'}(\vxi=0)=(2\pi\hbar)^{-1}N(\X')$.
We have left this normalization factor out of the definition,
because the basic interest is in the description of the full
state in the higher dimensional phase space. 
Indeed, for bound states there will be whole ranges of the parameters $\X'$
for which the overlap of the partial coherent state 
with $|\psi\rangle$ will be negligible. It will be verified in the
following sections that this occurs where the classical plane,
$\x'=\X'$, is not even close to intersecting the classical support of
$|\psi\rangle$.

So far we have not considered the possibility of
squeezing the coherent states $|\X'\rangle$, 
which generate the huwi function. In the limit
of infinite squeezing, these will be replaced 
by the position states $|Q'\rangle$, so that
the equation (\ref{sectwave}) 
for the wave function in the factor space becomes simply:
\be
\langle Q'|\psi\rangle(q_1)= \int {\rm d}q' \;\langle q'|Q'\rangle^*
 \;\langle q_1,q'|\psi\rangle = \langle q_1,Q'|\psi\rangle.
\label{redwave}
\ee
The corresponding Wigner function is just the limitting form of the
huwi function:
\be
h\!\!w_{Q'}(\x_1)=  \int {\rm d}\x'\; 
 W(\x_1, \x')\;\; \delta (q'-Q').
\ee
Hence, in this limit, the factor state results
from a combined momentum projection of the $p'$, 
with a (thin) section of the positions, $q'$.
In the case of a 4-D phase space, only one variable 
$q'=Q'$ is fixed, so this has become a 3-D section.

\section{Superpositions of coherent states}

\begin{figure}[t]
\begin{center}
\includegraphics[angle=-90,width=10.0cm]{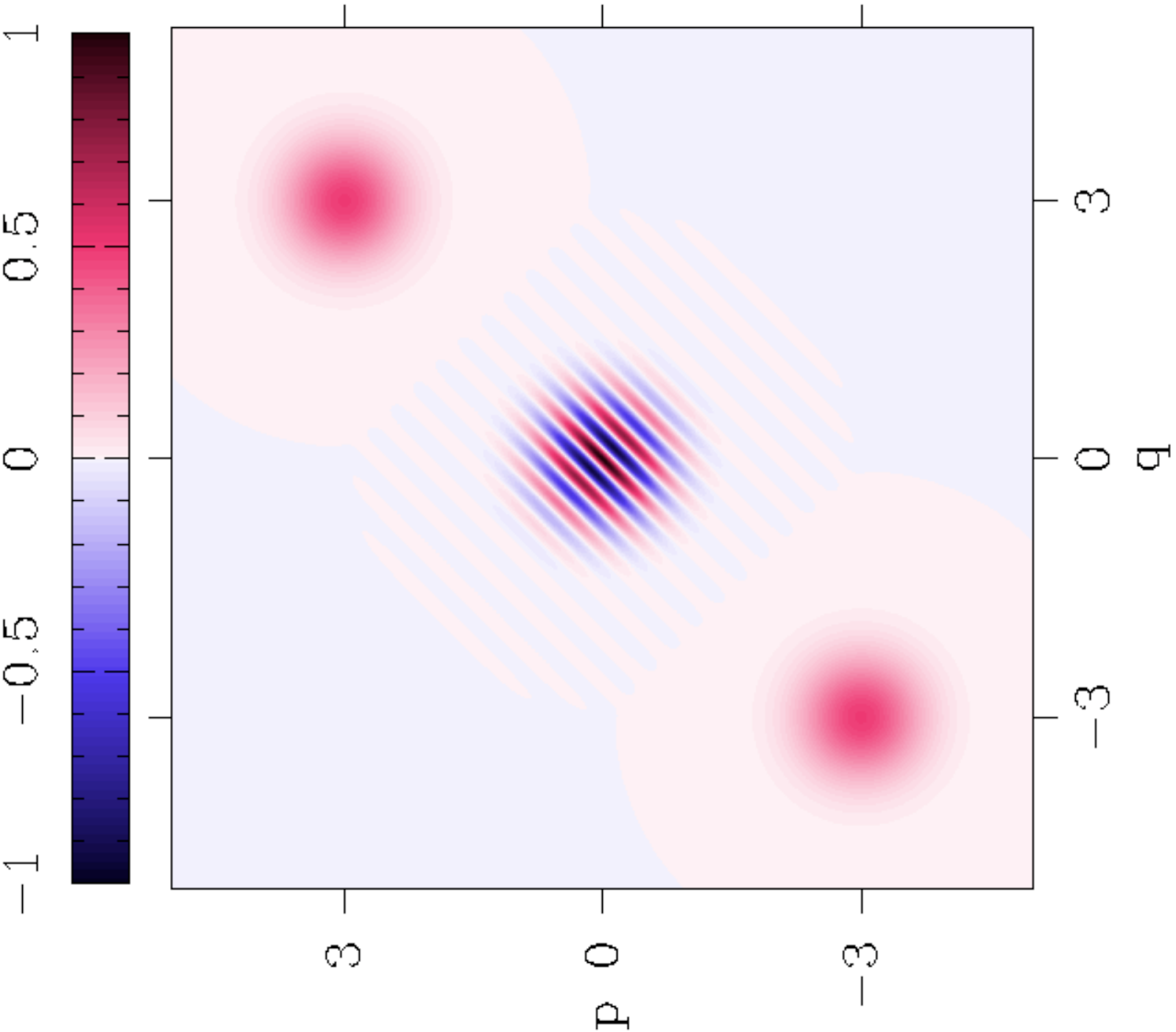}
\caption{The Wigner function for the Schr\"{o}dinger cat state 
displays a pair of {\it classical} Gaussians, one for each coherent state,
and a third Gaussian modulated by interference fringes halfway between them.
The chord function is a mere rescaling of the Wigner function if the midpoint
lies on the origin.}
\label{cat}
\end{center}
\end{figure}

It is well known that the coherent states (\ref{coherent})
have Gaussian Wigner functions (Schleich 2001, Ozorio de Almeida 1998):
\begin{equation}
\label{wcoherent}
 W_{\X}(\x) = \frac{1}{(\pi \hbar)^L} e^{-(\x-\X)^2/\hbar} \; ,
\end{equation}
whereas a superposition of coherent states, 
$|\X_a \rangle \pm |\X_b \rangle$, 
sometimes known as a {\it Schr\"{o}dinger cat state}, 
has the Wigner function: 
\begin{eqnarray}
W_{\pm}(\x) &=& 
\frac{1}{[2\pi \hbar \,(1\pm e^{-(\X_a - \X_b)^2/\hbar})]^L}
\left[ e^{-(\x-\X_a )^2 / \hbar } +   e^{-(\x-\X_b )^2 / \hbar } \pm
\right.\nonumber \\
&&\left.
\pm  2\; e^{-(\x-(\X_a + \X_b)/2)^2 / \hbar}
 \cos \frac{1}{\hbar} \x\wedge (\X_a-\X_b) \right].
\label{Wigcat}
\end{eqnarray}
It consists of two {\it classical} Gaussians centred on $\X_a$ and $\X_b$,
together with an interference pattern with a Gaussian envelope 
centred on their midpoint, as shown in figure \ref{cat}.  
The spatial frequency of this oscillation increases with the separation 
$|(\X_a-\X_b)|$. Increasing the number of coherent states merely increases the
number of classical Gaussians and adds new localized interference patterns
midway between each pair.

%

The overall picture does not depend on $L$, 
the number of degrees of freedom. Suppose then that $L$
is large and that we study classical 2-D sections, $\x'=\X'$,
of the Wigner function of a superposition of a
pair of coherent states that are centred on arbitrary points $\X_j$.
Clearly, $W(\x_1, \X')$ will only be appreciable if the chosen
$\X'$-plane is close to one of the ${\X_j}'$-planes 
on which the coherent states lie, or close to $(\X'_j+\X'_k)/2$, 
one of their midpoints, which houses the interference pattern. 
There will be no doubt about the localization of a coherent state
which is captured by a section close to ${\X_j}'$, 
but a section passing near a midpoint will not determine
where the interference pattern is coming from. 
Indeed, the spatial frequency of the oscillations within the section 
will depend only on the projection $(\X_j-\X_k)_1$, 
leaving $({\X_j}'-{\X_k}')$ completely undetermined.

There is little change between a classical section near an isolated coherent state
and a quantum section, i.e. a huwi function. Indeed, integrating over the product
of Gaussians in (\ref{huwi1}) produces a new Gaussian in the $\x_1$ plane, 
centred on $\X_{a1}$ or $\X_{b1}$.
However, if $\X'$ lies near $({\X_a}'+{\X_b}')/2$ the interference term is dampened by
a factor $\exp[-({\X_a}'-{\X_b}')^2/\hbar]$, 
so that it is not visible unless both section planes, 
${\X_a}'$ and ${\X_b}'$ lie close to each other.
Of course, this cancelling of interference is a familiar feature
of Husimi functions and merely reflects the Husimi side of 
the hybrid huwi representation. The Wigner side arises if
${\X_a}'$ and ${\X_b}'$ are nearly the same: Then there will
be practically no cancellation of the interference pattern
in the $\x_1$ plane, i.e. the huwi function will resemble the classical
section of the Wigner function, with two classical maxima 
and a central interference pattern. 

It might seem too severe a restriction only to observe interferences
of structures that lie on 2-D sections, but, because
the Wigner function is symplectically invariant, it is always possible
to picture the interference pattern between a pair of coherent states by effecting
a symplectic transformation which includes them both in the same 2-D plane.
On the other hand, the overlapping interference pattern for the superposition 
of a large number of coherent states will be vastly simplified in the huwi representation,
without reaching the extreme of the Husimi fuction itself.

In short, a superposition of coherent states is such a simple system 
that it would not be worth the bother of defining the huwi function to
deal with it. The purpose of this section is really to familiarize the reader
with an elementary example where we already have an adequate picture.
One can proceed to a more detailed investigation 
of superpositions of arbitrarily squeezed and rotated coherent states,
but it all reduces to performing Gaussian integrals 
and the qualitative result is unchanged:
The huwi representation displays the classical structure
cut by a given 2-D plane and the interfernce patterns due to this 2-D
structure, while washing out the interference effects of all classical structure
which has not been sampled by this plane.

\section{Van Vleck states}

A Van Vleck wave function $\langle q| \psi\rangle$ in higher dimensions 
(Van Vleck 1999 and  Ozorio de Almeida  1988), is 
suported by an $L$-D {\it Lagrangian} $\psi$-surface 
in the $(2L)$-D phase space (Arnold 1978). 
Locally, each branch of this surface is obtained by the action, 
\be
p^j(q)=\frac{\partial S^j(q)}{\partial q},
\label{lagder}
\ee
and the full wave function is then a superposition of the various branches:
\begin{eqnarray}
\langle q|\psi\rangle \approx \sum_j a^j(q)\> 
\exp \left [{
\over\hbar} S^j(q) \right ]. 
\label{wkbstate}
\end{eqnarray}
The amplitudes, $a^j(q)$, are determined by the classical structure 
and it suffices here to recall that they are finite except at
caustics (generalized turning points), 
where the branches of the function $S^j(q)$ are joined.
The semiclassical approximation to the wave function breaks down
at caustics, but these can be shifted by symplectic transformations.
In the case of a bound state, 
the $\psi$-surface is an $L$-D torus (Arnold 1978). 
All closed curves on the $\psi$-surface
must satisfy approximate Bohr-Sommerfeld quantization rules,
that is, 
\be
\oint_{\psi_1}p_1\> 
q_1=(n+{1\over 2})\hbar.
\label{BSq}
\ee
Thus the Van Vleck state is a generalization of simple WKB states.

The semiclassical form of the Wigner function for generalized WKB states
was derived by Berry 1977a, but, rather than obtaining the huwi function 
by its integration with a Gaussian window,
it is simpler to start from (\ref{sectwave}) and (\ref{wkbstate}) 
to calculate the reduced wave function:
\begin{eqnarray}
\langle X'|\psi\rangle(q_1) \approx \sum_j  \int {\rm d}q'\; 
\frac{a^j \!(q_1, q')}{(\pi \hbar)^{(L-1)/4}}\> &&
\exp \left [-\frac{1}{2\hbar}(q'-Q')^2-\frac{i}{\hbar}P' \cdot(q'-\frac{Q'}{2})+
\right. \nonumber \\
&&\left.+\frac{1}{\hbar} S^j(q_1, q') \right ].
\label{overlap1}
\end{eqnarray}
The Gaussian factor in the integrand allows us to expand the action
around $Q'$ as
\be
S^j(q_1, q') \approx S^j(q_1, Q') + {p^j}'(q_1, Q') \cdot (q'-Q'),
\ee
where 
\be
{p^j}'(q_1, Q')=\frac{\partial S^j}{ \partial q'}.(q_1, q'=Q')
\ee
Then (\ref{overlap1}) reduces to the Fourier integral of a Gaussian,
if the slow variation of $a^j \!(q_1, q')$ is neglected, that is, 
\begin{eqnarray}
 \langle X'|\psi\rangle(q_1) \approx \sum_j   
(4\pi \hbar)^{(L-1)/4} \; a^j \!(q_1, Q') \nonumber \\
\times\exp \left [-\frac{1}{2\hbar}({p^j}'(q_1, Q')-P')^2
-\frac{i}{2\hbar}P' \cdot Q'
+{
\over\hbar} S^j(q_1, Q') \right ].
\end{eqnarray}
This wave function is localized in the neighbourhood of the point ${Q_1}^j$,
defined by the equation, ${p^j}'({Q_1}^j, Q')=P'$, that is, the position that
corresponds to the intersection of the $X'$-plane 
with the $\psi$-surface .
A linear expansion of the action, $S^j(q_1, Q')$, around this point
then leads to the simplified expression:
\begin{eqnarray}
\langle X'|\psi\rangle(q_1) \approx \sum_j   
(4\pi \hbar)^{(L-1)/4}\; a^j \!({Q_1}^j, Q') \; 
\exp \left[-\frac{1}{2\hbar}P' \cdot Q' +{
\over\hbar}  S^j({Q_1}^j, Q')\right]  \nonumber \\ 
\times\exp \left[-\frac{1}{2\hbar}(q_1 - Q_1^j)\cdot{\bf \Omega}^j\cdot(q_1 - Q_1^j)
  +{
  \over\hbar}p_1(Q_1^j) \cdot(q_1 - Q_1^j) \right],
\end{eqnarray}
where
\be
{\bf \Omega}^j({q_1}^j, Q')=\frac{\partial S^j}{\partial q_1 \partial q'} ({q_1}^j, Q')
\ee
is the frequency matrix for a generalized squeezed and rotated coherent state. 

Thus, the reduced wave function obtained from a Van Vleck state
by a quantum section is approximately 
a superposition of generalized coherent states, 
if the section plane intersects the $\psi$-surface 
at isolated points. This is the generic situation, 
because the $\X'$-plane is 2-D and the $\psi$-surface is $L$-D.
Thus, for $L=2$, we have generic point intersections for continuous regions of $\X'$,
whereas, for $L>2$, the section plane may easily miss the $\psi$-surface,
but, where intersections occur, they are generically at isolated points.
The situation is quite analogous to the intersections of a given curve
with a set of parallel lines in 3-D.

The huwi representation,
can then genererically be approximated by a Schr\"odinger cat state,
with the generalized coherent states placed on the $\x_1$-projection 
of the points of intersection of the $\psi$-surface with the section plane.
This is similar to the states considered in the previous section. 
The main difference is that, for $L=2$, variations of the $\X'$-section plane 
lead to similar patterns in the case of Van Vleck states, 
because these also intersect the $\psi$-surface. 
In contrast, the huwi representation of cat states 
changes drastically depending on whether the section plane 
comes close to any of the localized states within the full phase space.
Also, it will be unusual for a higher dimensional cat to have more than
one coherent state sampled by a given $\X'$-section, 
whereas the general huwi representation of a Van Vleck state, for $L=2$, 
is a full cat with its interference patterns at all the midpoints.  

\begin{figure}[t]
\begin{center}
\includegraphics[width=10.0cm]{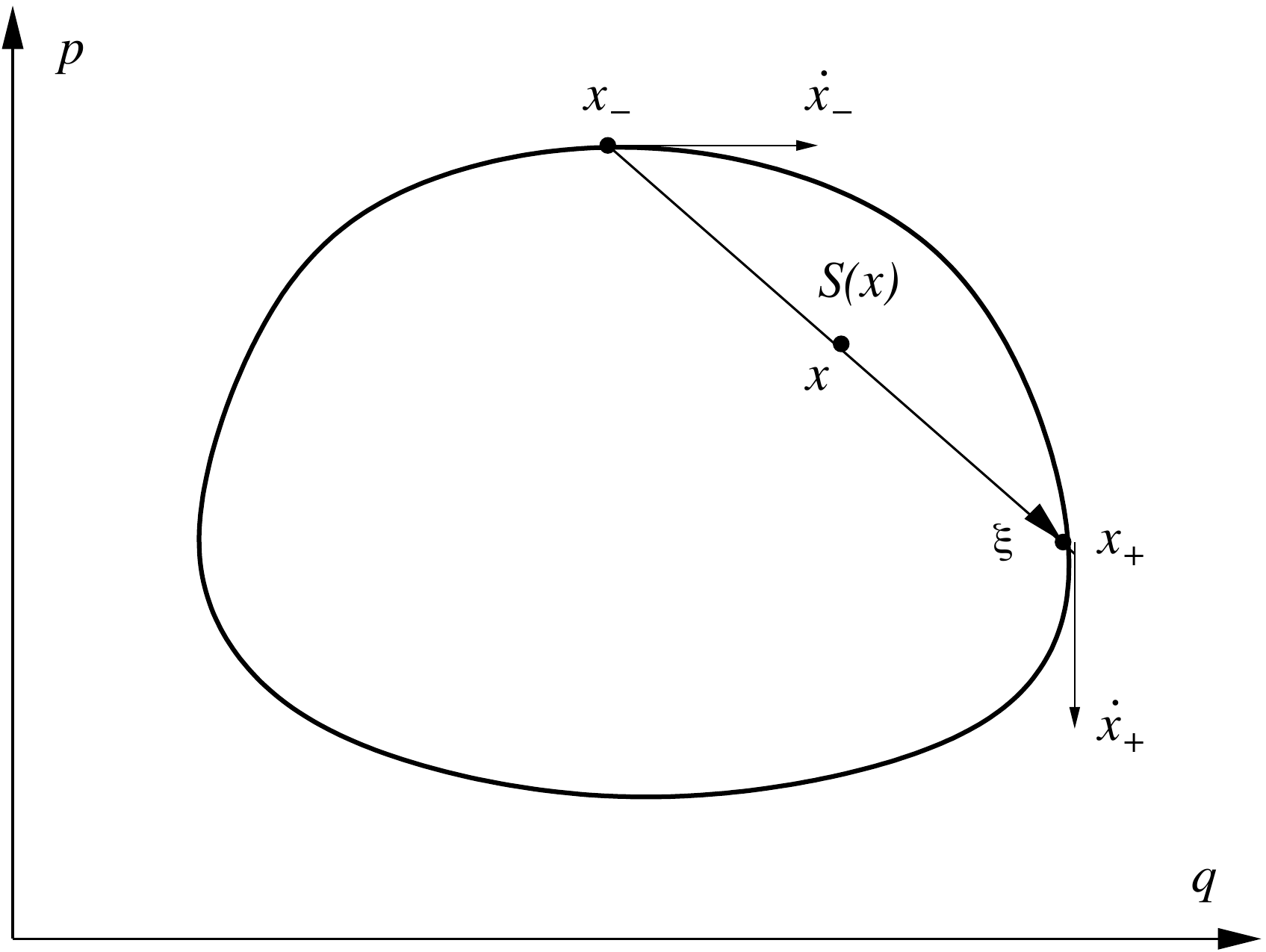}
\end{center}
\caption{A single chord, $\vxi$, is centred on $\x$ if it is close to a convex quantized curve
in the case of one degree of freedom. The phase of the Wigner function is proportional
to the area $S(\x)$ between the chord and the shell, while the amplitude, $A(\x)$, depends
on both phase space velocities at the tips of $\vxi$. A caustic results from parallel
tangents at the chord tips. }
\label{fig2}
\end{figure}

It is now necessary to discuss the nongeneric case
where the intersection of the $\X'$-plane and the $\psi$-surface 
is not a set of isolated points: Consider the case of a product 
state, $|\psi_1\rangle \otimes|\psi'\rangle$ , 
then its classical support will be a Lagrangian surface
that is a cartesian product of lower dimensional surfaces, 
$\psi_1$-surface$\times\psi'$-surface. In the case of a bound 
state, the $\psi$-surface is thus factored as 
a $(L-1)$-D torus and a closed $\psi_1$-curve.
Then any nonempty intersection of the $\X'$-plane with the $\psi$-surface
produces the full $\psi_1$-curve (Ozorio de Almeida \& Hannay 1982). 
All closed curves on the $\psi$-surface
must satisfy approximate Bohr-Sommerfeld quantization rules (\ref{BSq}),
so that the $\X'$-section is also a quantized curve. Thus, the huwi representation of this product state 
is approximately the semiclassical Wigner function for $L=1$  (Berry 1977a):
\begin{equation}
W_1(\x_1)\approx\sum_k A^k(\x_1) \> \cos{S^k(\x_1)\over\hbar}.
\label{decompure}
\end{equation}
The sum in \ref{decompure} runs over all {\it chords} 
on the $\psi_1$-curve that are centred on $\x_1$ 
and $S_k$ is the area between the chord and the shell 
(plus a semiclassically small {\it Maslov phase}) as shown in figure \ref{fig2}.  
The semiclassical approximation \ref{decompure}
breaks down along {\it caustics}, where the amplitudes
$A^k$, display spurious divergences. The caustics of Wigner functions are the loci
of coalescing chords.

%

\begin{figure}[t]
\begin{center}
\includegraphics[ width=12.0cm]{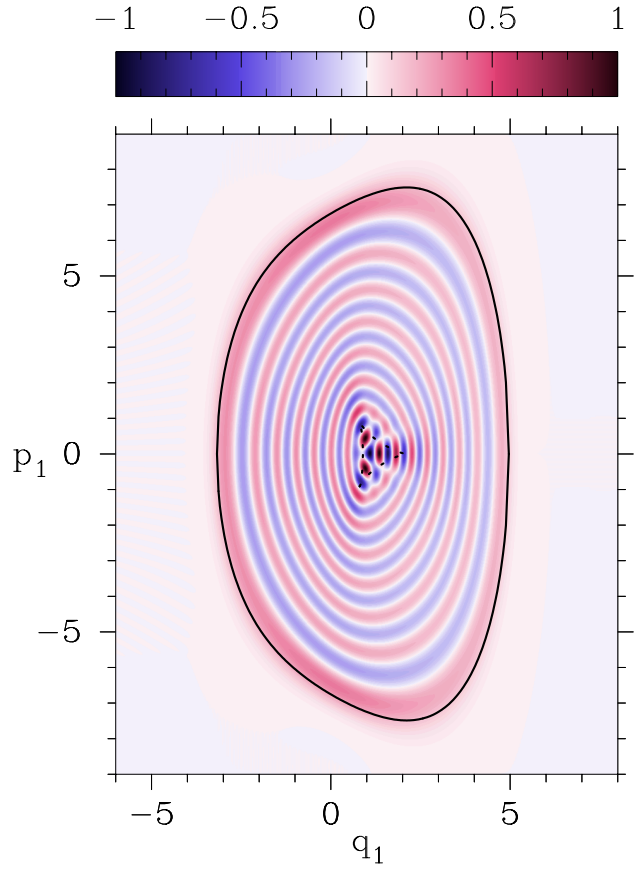}
\end{center}
\caption{The Wigner function of a $L=1$ degrees of freedom system 
associated with a quantized $\psi_1$-curve
(the full line). Inside the Wigner caustic (the dotted line) the interference pattern
comes from the presence of three chords for each phase space point $\x_1\equiv(p_1,q_1)$.}
\label{fig3}
\end{figure}

Figure \ref{fig3} displays an exact Wigner function for $L=1$, where the semiclassical
structure is clearly discernible. The quantized $\psi_1$-curve is a uniform maximum
with a constant phase and within it there lie a succession of constant phase rings.
In this region there is only one chord, $\vxi(\x)$. 
In the central region, the pattern becomes more complex, 
owing to the presence of three chords for each phase space point.
It is remarkable that the relation of the interference pattern, 
at any point $\x$ in the interior of the curve, 
to the chord, $\vxi(\x)$, is quite analogous
to that between the interference fringes of a Schr\"odinger cat and the
vector that separates the pair of coherent states:
the analogy holds for the direction of the fringes and their spacial frequency.
Indeed, it is possible to fit the semiclassical state quite accurately 
by a generalized cat state with a discrete set of
coherent states along the quantized curve (Kenfack  \textit{et al.} 2004 and
Carvalho \textit{et al.}  2006).
Evidently, the phase difference between the pair of coherent states
at the chord tips must then agree with $S(\x)$, the area between the chord 
and the quantized curve.

Even tori which are not products will produce huwi functions
in the standard form (\ref{decompure}) if this is generated by
a position state, considered as the limit of the squeezed states
analyzed at the end of section 2. Evidently, fixing $Q'$
in (\ref{redwave}) generates the semiclassical wave function,
\begin{eqnarray}
\langle Q'|\psi\rangle(q_1)=
\langle q_1, Q'|\psi\rangle \approx \sum_j a^j(q_1, Q')\> 
\exp \left [{
\over\hbar} S^j(q_1, Q') \right ],
\end{eqnarray}
which is also of WKB form. Moreover, all the different $(2L-1)$-D
sections, defined by each $Q'$, 
must intersect the torus along quantized closed curves,
i.e. satisfying (\ref{BSq}). Therefore, each of the
corresponding Wigner functions, $h\!\!w_{Q'}(\x_1)$, 
has the usual form (\ref{decompure}).

In the following section, we display computational evidence that
the huwi function of ergodic quantum states resemble in some aspects 
semiclassical Wigner functions of bound states of simple systems
with a single degree of freedom.

\section{Numerical study of eigenstates of a chaotic system}

The intersection of the 2-D $\X'$-plane with a 
$(2L-1)$-D compact energy shell of a bound classical system
produces a closed curve in the $\x_1$-plane, for all $L>1$.
Therefore, it is the interference of this classical structure
within the section that should determine the huwi representation
for each parameter $\X'$. Viewed as a pure state for $L=1$,
this is a type of Wigner function that has not been previously studied.
A resemblence to the simple semiclassical
Wigner function presented at the end of the previous section may be anticipated:
the chords, $\vxi(\x)$, on the section curve should lead to interference fringes
parallel to the chord, just as in the WKB case. The room for freedom
lies in the way this phase is shifted, as the centre $\x$ is displaced.
Of course, our semiclassical intuition cannot be pushed too far.
The huwi function results from a {\it thick} quantum section, 
rather than a {\it thin} classical section, 
so that a neighbourhood of the full Wigner function is involved.
Furthermore, we must deal with a continuum of $\X'$-parametrized
Wigner functions, instead of a discrete set of Bohr-quantized states.

\begin{figure}[t]
\includegraphics[angle=-90,width=20cm]{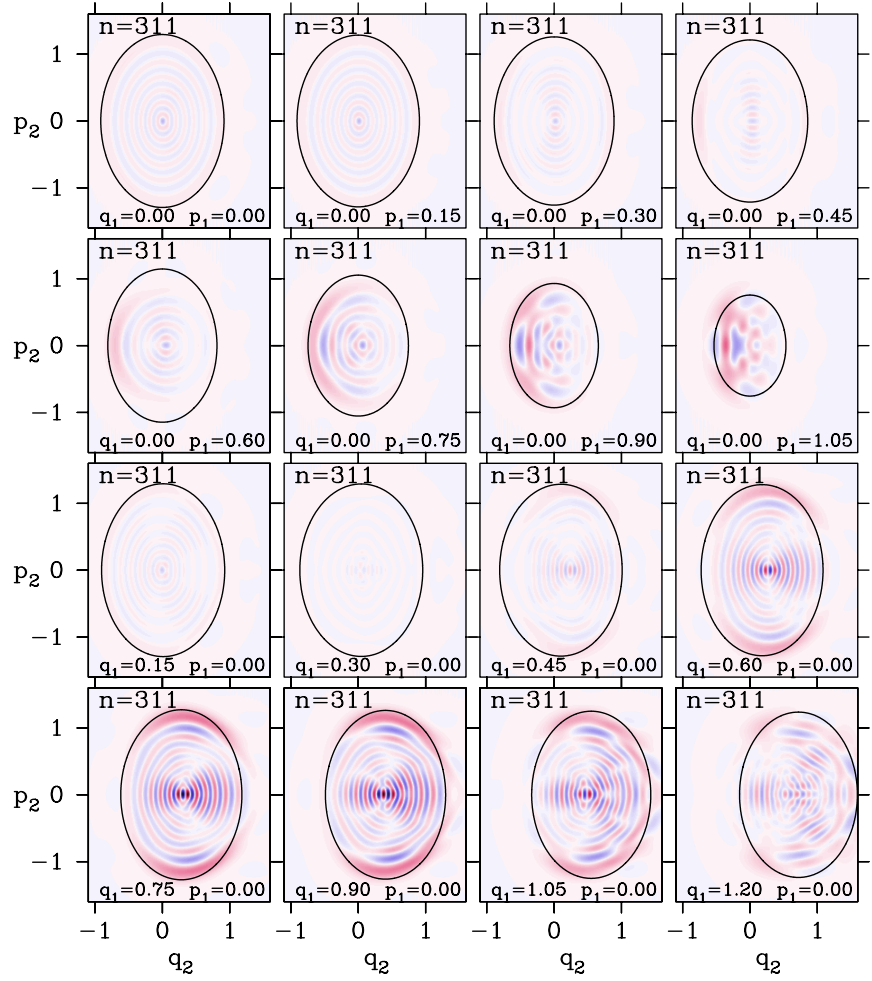}
\caption{The huwi functions for the eigenstate, 
$|\psi_{n=311}\rangle$, of the {\it Nelson} 
Hamiltonian in different 2-D planes  $q_1=Q'$ and $p_1=P'$. 
The black closed curve in each graph is the intersection of the energy shell 
with the specified 2D-plane for the eigenenergy $E_{n=311}$. 
This is the classical trajectory of the harmonic oscillator Hamiltonian 
obtained by restricting the original Hamiltonian to that plane.  
Relative intensities of the huwi functions are displayed for all the graphs.}
\label{fig4}
\end{figure}

In the absence of a full semiclassical theory for chaotic eigenstates
we resort to numerical calculations of the huwi function for a particular system.
The classical {\it Nelson} Hamiltonian is defined as,
\be
\label{hamilnelson}
\mbox{H}({\bf x}_1,{\bf x}_2)=
(p_1^2+p_2^2)/2 + 0.05\,q_1^2+
(q_2-q_1^2/2)^2
\;,
\ee 
and its quantum counterpart results from the replacement: 
${\bf x}_1\equiv(q_1,p_1)\rightarrow (\hat{q_1},\hat{p_1})$ and
${\bf x}_2\equiv(q_2,p_2)\rightarrow (\hat{q_2},\hat{p_2})$ in (\ref{hamilnelson}).
The restriction of the  classical Hamiltonian (\ref{hamilnelson}) 
to the 2-D $\X'$-planes $q_1=Q^{'}$ and $p_1=P^{'}$
defines harmonic oscillators in the $\x_2$ variables: 
$h_2(\x_2)=\mbox{H}(\X',{\bf x}_2)$. 
The classical trajectories of this harmonic oscillator
coincide with the intersection of the energy shell of the full Hamiltonian 
with the planes $q_1=Q^{'}$ and $p_1=P^{'}$.
The alternative classical sections $\x_2=\X'$ on (\ref{hamilnelson}) defines
anharmonic oscillators, $h_1(\x_1)=\mbox{H}({\bf x}_1,\X')$, 
whose closed trajectories are not ellipses.  

\begin{figure}[t]
\setlength{\unitlength}{1cm}
\begin{center}
\includegraphics[angle=-90,width=20.cm]{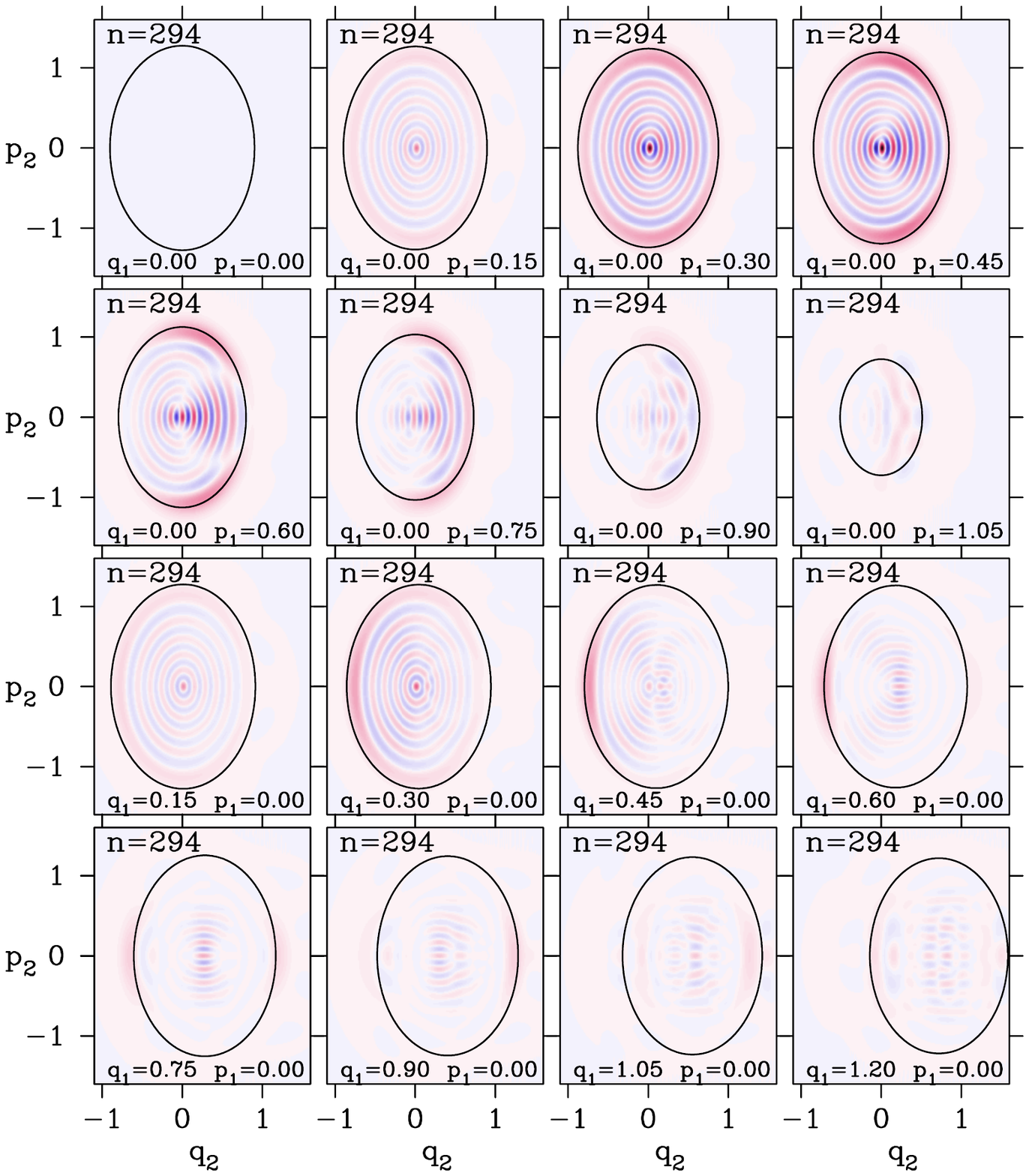}
\end{center}
\caption{Idem figure \ref{fig4}, but for the eigenstate 
the {\it Nelson} Hamiltonian. Note the the huwi function is null, in the symmetry plane 
${\bf x}_1=\X'=(0,0)$, because its definition involves the partial projection 
$\langle \X'|\psi_{n=294}\rangle$ of an even coherent state $|\X'\rangle$ 
with an odd $|\psi_{n=294}\rangle$ state in the ${\bf x}_1$ degree of freedom.}
\label{fig5}
\end{figure}

\begin{figure}[t]
\setlength{\unitlength}{1cm}
\begin{center}
\includegraphics[angle=-90,width=20.5cm]{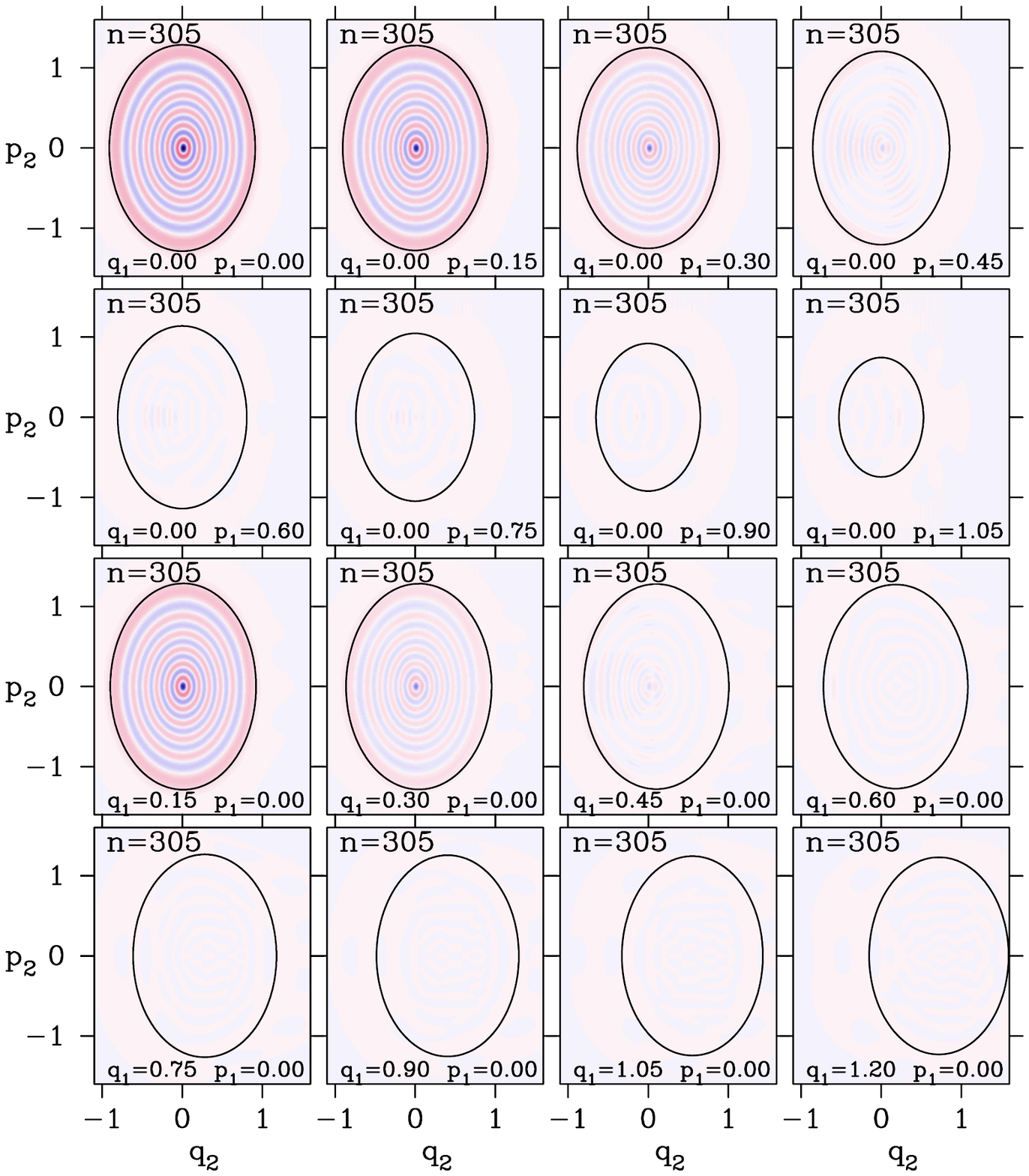}
\end{center}
\caption{Idem figure \ref{fig4}, but for the eigenstate 
The enhanced amplitude of the huwi functions around the classically invariant plane  
reflects the fact that the Wigner function of this state has a scar of 
the periodic orbit in this plane (see figure \ref{fig7}). 
The distance of the eigenenergy $E_{n=305}$ from the Bohr energy level 
for the periodic orbit is of the order of a single level spacing, i.e. ${\cal O}( \hbar^2)$. }
\label{fig6}
\end{figure}

The classical dynamics of this system is mixed for all energies, but we consider 
eigenstates $|\psi_n\rangle$ whose eigenenergies correspond to an essentially 
chaotic classical behavior (see Prado \& de Aguiar  1994 and references therein). 
It would be natural to use a harmonic oscillator basis 
for calculating those of the Nelson Hamiltonian,
but it is more efficient to use a basis of {\it distorted oscillators}
(Toscano \textit{et al.}  2001). In the computations we fixed $\hbar=0.05$.
In Figs.(\ref{fig4}),(\ref{fig5}) and (\ref{fig6}) 
we show the huwi function of three different eigenstates
for several  constant $\x_1$ planes. 
It should be observed that, in spite of varying degrees of irregularity 
in the internal fringe pattern among these examples, 
the wavelength is larger for the smaller classical curves,
which hence have shorter chords. 
In all cases the pattern decays outside the classical curve. 

\begin{figure}[t]
\setlength{\unitlength}{1cm}
\begin{center}
\includegraphics[angle=-90,width=20.cm]{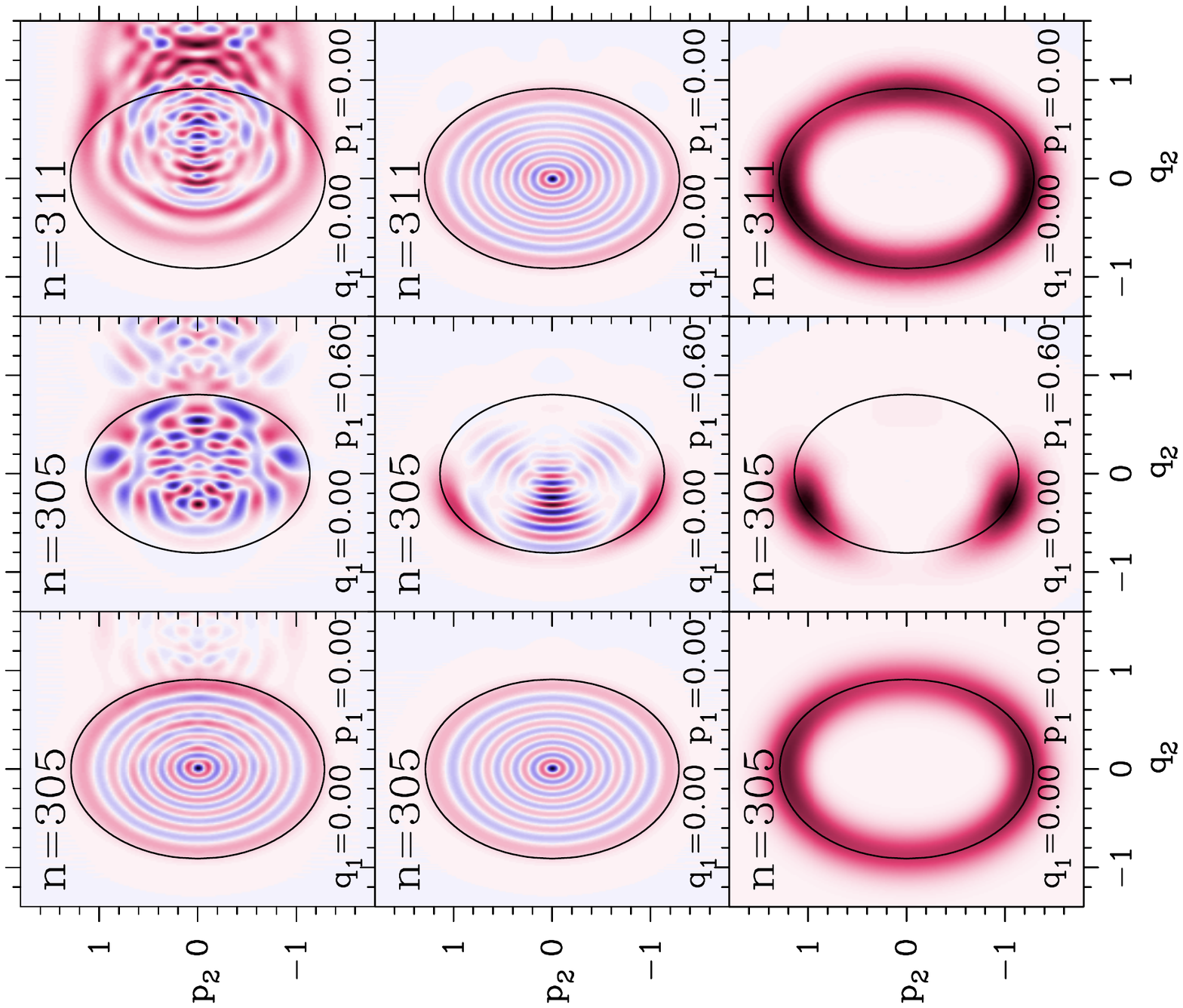}
\end{center}
\caption{The top row displays plane (thin) sections 
of the Wigner function for three eigenstates
$|\psi_n\rangle$ of the Hamiltonian (\ref{hamilnelson}). 
The classical curve is drawn in black,
that is the energy shell in each plane $q_1=Q'$ and $p_1=P'$.
These sections have interference contributions of different parts of 
the energy shell far from these plane sections (particularly evident  
in the region $q_2 < 0$ outside the classical curve).
The middle row shows the corresponding huwi functions and the bottom row
displays the sections of the Husimi functions.}
\label{fig7}
\end{figure}

It is interesting to compare the huwi functions 
to thin sections of the full Wigner function of the eigenstates,
along the planes $q_1=Q^{'}$ and $p_1=P^{'}$ as in figure \ref{fig7}, which
also compares them to the Husimi functions for the corresponding quantum sections. 
These sections of the full Wigner function 
display interference of contributions from different parts of 
the energy shell outside of the plane section. 
Semiclassically, these oscillations can be ascribed to chords,
$\xi(\x)$, with centers, $\x$, in the section, 
but with tips on the energy shell far from this plane.
This is particularly evident in the region
$q_2<0$ outside the classical curve.
The Gaussian smoothing (\ref{huwi1}), that defines the huwi function 
in the planes $q_1=Q^{'}$ and $p_1=P^{'}$, 
washed out all these interference contributions,
isolating the contributions from chords with tips 
lying on the classical curve, or in its neighbourhood. 
Thus, the result is a huwi function 
with an interference pattern inside the classical curve 
and exponentially small values outside, 
which resembles the Wigner functions associated with a quantized curve
(see figure \ref{fig2}).
However, it is clear that the constant phase curves of the interference fringes  
do not follow exactly the regular concentric pattern 
of a Wigner function for a quantized curve.

The plane $q_1=p_1=0$ is special for the Nelson Hamiltonian.
On the one hand, it is the symmetry plane 
for the reflection symmetry $q_1\rightarrow -q_1$ of the system. 
Thus, the huwi function in this plane discriminates the odd from the even states
in the ${\bf x}_1$ degree of freedom.
For odd states the huwi function is zero on the plane, 
because, according to (\ref{sandw}), it comes from a partial 
projection on an even coherent state $|\X'=(0,0)\rangle$.
On the other hand, this is a classically invariant plane 
that contains the {\it vertical family} of periodic orbits
of the full Hamiltonian.
All the chords with tips on one periodic orbit 
define the so called {\it central surface}, 
which, in this case, is merely the region in the invariant plane 
inside the classical vertical orbit (Toscano \textit{et al.}  2001). 
When the classical orbit is Bohr (or "anti-Bohr") quantized, 
it was shown in Toscano \textit{et al.}  2001 that the mixture of eigenstates in a narrow
energy window, {\it i.e.} the spectral Wigner function, 
present a scar in the central surface of the periodic orbit, 
which takes the form of a pattern of concentric rings
of constant phase, just like those of the semiclassical 
$L=1$ Wigner function (\ref{decompure}).
This pattern is also evident for an individual eigenstate, 
as is the case for $|\psi_{n=305}\rangle$, 
whose eigenenergy is very close to the Bohr energy (see figure \ref{fig7}).
In this case the huwi function in the invariant plane, 
as a thick section of the Wigner function, 
simply isolates the scar contribution of the periodic orbit (see figure \ref{fig6}).
The thickness of the scar is sampled by the huwi functions 
up to a linear distance of the order $\sqrt{\hbar}\approx 0.22$
by sections parallel to the symmetry plane.

Even when the periodic orbit is far from being Bohr quantized, the huwi functions
around the invariant plane isolates similar contributions from the periodic orbit, 
as we can see in figure \ref{fig4}  and  \ref{fig5}. 
The similarity of these huwi functions with the Wigner functions
for harmonic oscillator eigenstates is remarkable, 
specially if it is recalled that their energies are not Bohr-quantized.

The quantum sections which are represented by the huwi function, 
should not be confused
with the {\it Quantum Poincare Surface of Section} (Leboeuf  \& Saraceno 1990a  
and Leboeuf  \& Saraceno 1990b). 
The latter is a {\it projection} onto a 2-D plane 
of the Husimi function evaluated along the 3-D energy shell. 
It is an invaluable tool for the study of the classical features
of the quantum state. For instance, scarred states (Heller 1984,  Bogomolny 1988 \& Berry  1989 ) 
exhibit maxima in the neighbourhood of the points where a periodic orbit 
crosses the corresponding classical section (see e.g  Arranz \textit{et al.}  2004). 
In contrast, the 2-D quantum section 
corresponds to a classical section that intersects the energy shell 
along a closed curve. The Husimi representation of this quantum section
coincides with the section of the Husimi function (\ref{Hsection})
and, hence, it is entirely concentrated along the classical curve.
Thus, the structure within the classical curve of the huwi function
is built up of quantum interferences which are washed out
in the Husimi representation of the section, as shown in figure \ref{fig7}.

\section{Discussion}

The huwi representation displays the classical structure
cut by a given 2-D plane and the interfernce patterns due to this 2-D
structure, while washing out the interference effects 
of all classical structure, which has not been sampled by this plane. 
This provides a valuable tool for the study of the eigenstates
of chaotic systems, whether these be ergodic, or scarred to some extent.
 
The Nelson Hamiltonian, whose eigenstates were presented in the previous section,
corresponds to a mixed classical system. At low energies, the
motion is a perturbation of the 2-D harmonic oscillator, 
but becomes increasingly chaotic at higher energies.
Even though the limit of truly ergodic motion is unlikely ever to be reached,
it is expected that most orbits will eventually sweep over most of the energy
shell, within the wavelength of the quantum motion for the states that we have
considered. Therefore, we expect that most eigenstates will resemble 
ergodic quantum states. 

Our computations clearly show that even a partial
smoothing of the Wigner function is not concentrated in the neighbourhood
of the energy shell. Indeed, each quantum section of the energy shell,
sampled by each huwi function, marks the boundary, 
beyond which the huwi function is vanishingly small. 
But inside each section of the shell, a closed curve, 
the huwi function oscillates somewhat like a typical Wigner function of a
Bohr-quantized state in the case of a single degree of freedom.
Usually these are considered as simple examples of eigenstates
of integrable systems, though it should not be forgotten
that they are also trivially ergodic: 
the trajectories visit uniformly the entire energy shell.
Therefore, the huwi representation has brought to light a natural, but unsuspected
family resemblence between ergodic states of all dimensions. 

The difference between huwi functions for chaotic eigenstates and
simple Bohr-quantized states concerns the regularity of the wave
pattern inside the classical curve. With the exception of 
huwi functions obtained from very symmetric sections of the energy shell,
it was found in the previous section that the regular pattern of
concentric wave fronts for the bohr-quantized state in figure \ref{fig2} 
may be broken up in many places, even though the direction of the
phase curves and their spacing is approximately maintained.

This scenario is curiously reminiscent of a snap shot of
dislocations in wave trains, analyzed by Nye \& Berry 1974.  
For travelling wave trains, the approximately constant wavelength
is a consequence of the approximately constant temporal frequency of 
an initial pulse. However, the returning signal 
results from several scattered components with different phases.
The outcome is an imperfect wave pulse with dislocations, where
constant phase curves are interrupted, just as for an imperfect
crystal lattice. Similar structures arise for travelling waves 
through spatial temporal disorder (La Porta \&  Surko  1996) . 
In the present case, it is the single geometric chord
centred on a given interior point of the classical curve
that specifies the dominant spacial frequency near the centre, 
as well as the direction of the wave fronts, which must be parallel to the chord. 
So far, this is the same as holds for the Bohr-quantized curve,
but the neighbouring oscillations of the chaotic huwi functions 
are closer to scattered wave trains then to regular concentric waves.
This same general picture can be conjectured for 2-D quantum sections
of even higher dimensional chaotic energy shells. 

Perhaps, this is only a very qualitative analogy, though
it indicates a direction to be pursued for the systematic characterization
of huwi functions obtained from chaotic eigenstates.
It is emphasised in Nye \& Berry 1974 that the finite size of the
pulse is a precondition for the presence of dislocations.
In this respect, it should be recalled that diverse types of quantum states
can be successfully fitted by localized coherent states placed along the
relevant classical manifold, such that the oscillations 
midway between each pair of phase space Gaussians has the same spatial
frequency as that of the fitted state (Kenfack  \textit{et al.} 2004 and 
Carvalho \textit{et al.}  2006).
One can then conjecture that the dislocations 
on the crests of the huwi pattern may be ascribed
to the superposition of neighbouring coherent states on the classical curve 
with arbitrary phases.

\begin{acknowledgements}

Partial financial support from 
Millenium Institute of Quantum Information, PROSUL, FAPERJ and CNPq is gratefully 
acknowledged. AK acknowledges the financial supports by the Max-Planck 
Gessellschaft through Reimar L\"ust fund (Reimar L\"ust Fellow 2005) 
and by the Alexander von Humboldt foundation with the research grant 
No.IV. 4-KAM 1068533 STP.

\end{acknowledgements}

\end{document}